\documentclass[%
 reprint,
 amsmath,amssymb,
 aps,
]{revtex4-1}

\usepackage{graphicx}
\usepackage{dcolumn}
\usepackage{bm}

\usepackage{subfig}
\usepackage{xcolor}
\usepackage{subfig}
\usepackage{amsmath}
\usepackage{graphicx}
\usepackage{epstopdf, epsfig}
\usepackage{bm}
\usepackage{color,soul}
\usepackage{textcomp}

\begin{document}

\preprint{APS/123-QED}

\title{Droplet Splashing on Rough Surfaces}

\author{Thijs de Goede}
\affiliation{%
 van der Waals-Zeeman Institute, Institute of Physics, University of Amsterdam, Science Park 904, 1098XH Amsterdam, the Netherlands
}%

\author{Karla de Bruin}
\affiliation{%
 van der Waals-Zeeman Institute, Institute of Physics, University of Amsterdam, Science Park 904, 1098XH Amsterdam, the Netherlands
}

\author{Noushine Shahidzadeh}
\affiliation{%
 van der Waals-Zeeman Institute, Institute of Physics, University of Amsterdam, Science Park 904, 1098XH Amsterdam, the Netherlands
}

\author{Daniel Bonn}
 \email{D.Bonn@uva.nl}
\affiliation{%
 van der Waals-Zeeman Institute, Institute of Physics, University of Amsterdam, Science Park 904, 1098XH Amsterdam, the Netherlands
}%

\date{\today}

\begin{abstract}
When a droplet hits a surface fast enough, droplet splashing can occur: smaller secondary droplets detach from the main droplet during impact. While droplet splashing on smooth surfaces is by now well understood, the surface roughness also affects at which impact velocity a droplet splashes. In this study, the influence of the surface roughness on droplet splashing is investigated. By changing the root mean square roughness of the impacted surface, we show  that the droplet splashing velocity is only affected when the droplet roughness is large enough to disrupt the spreading droplet lamella and change the droplet splashing mechanism from corona to prompt splashing. Finally, using Weber and Ohnesorge number scaling models, we also show that the measured splashing velocity for both water and ethanol on surfaces with different roughness and water-ethanol mixtures collapse onto a single curve, showing that the droplet splashing velocity on rough surfaces scales with the Ohnesorge number defined with the surface roughness length scale.

\end{abstract}

\maketitle


\section{Introduction}

Droplet impact upon solid surfaces is a phenomenon of importance to a wide range of applications. If the velocity with which a droplet impinges on a surface is above a critical value called the splashing velocity, the droplet breaks up after impact. Numerous studies on droplet splashing have shown that increasing the surface roughness significantly reduces the splashing velocity  \cite{Stow1977,Range1998,Rioboo2001,Wal2006,Latka2012,Roisman2015,Hao2017,Quetzeri2019wetrough}. However, a thorough investigation of the relation between surface roughness and splashing is complicated by the many parameters of the problem, and by the characterization of the surface roughness. Different definitions of the roughness are in fact possible, some authors use either the arithmetic ($R_a$) or root mean square ($R_{rms}$) roughness of the surface \cite{Range1998,Latka2012}, while others have argued that splashing depends on a characteristic slope of the surface profile \cite{Roisman2015}, or account for the multiscale roughness of the surface \cite{Quetzeri2019wetrough}. In this study, we will investigate surfaces with a roughness characterized by only a single length scale, which we will characterise with the root mean square roughness of the surface.
\\   

In addition, recent studies have shown that wettability also influences droplet splashing \cite{Quetzeri2019,Quetzeri2019wetrough}. In contrast to earlier reports \cite{deGoede2017,Latka2018}, these studies show that for low wettability surfaces (contact angles $\theta > 110^\circ$), the splashing velocity is lower than for wettable surfaces ($\theta < 90^\circ$). This effect of wettability is important for droplet splashing on rough surfaces as the surface wettability is linked to the surface roughness, especially for low wettability and non-wetting surfaces \cite{Gennes2004,Bonn2009}.\\

In this study the influence of the surface roughness on droplet splashing is investigated for Newtonian fluids. By experimentally measuring the splashing velocity for surfaces with a wide range of root mean square roughness, we observe a transition in splashing mechanism when the surface roughness crosses a certain threshold. This threshold is found to also depend on the wetting properties of the liquid on the solid. This was varied both by using water-ethanol mixtures to tune the surface tension of the fluid, and different solids to tune the surface tension of the solid. We find that below a certain surface roughness, the splashing velocity is independent of surface roughness and can be described by the splashing model of Riboux and Gordillo for smooth surfaces\cite{Riboux2014}. Above a critical roughness, the splashing velocity decreases with increasing roughness. We show that a critical Weber scaling model proposed by Garc{\'i}a-Geijo  \textit{et al.} \cite{Quintero2020} adequately describes the relation between the splashing velocity and surface roughness. The Weber number represents the balance of inertial to capillary forces. This scaling model therefore does not take the influence of viscosity into account, which we find changes the splashing threshold somewhat. We therefore investigate a heuristic scaling model based on the Ohnesorge number of the fluid at the length scale of the surface roughness. We find that a $Oh^{2/3}$ scaling model works very well in collapsing the data, indicating that the inertial, viscous and capillary forces are all important for (prompt) splashing,. However we cannot provide a theoretical derivation of this scaling, leaving the question of the viscous contribution to the splashing threshold open. 
      
\section{Material and Methods}
\begin{table}[htb]
\centering
\caption{Water, ethanol, and ethanol-water mixtures fluid parameters, sorted by ethanol weight
percentage (wt \%). Sources: Refs. \cite{Haynes2014,Vazquez1995}.}
\renewcommand{\arraystretch}{2}
\centerline{
  \begin{tabular}{ c | c | c | c }
wt \%& Density & Surface Tension & Viscosity
\\
ethanol & $\left( \frac{\mathrm{kg}}{\mathrm{m}^3} \right)$ & $\left( \frac{\mathrm{mN}}{\mathrm{m}} \right)$ & 
$\left( \mathrm{mPas} \right)$\\
  \hline
  0 & 997.0 & 71.99  & 0.89 \\
  \hline
  5 & 989.0 & 56.41  & 1.228 \\
  \hline
  10 & 981.9 & 48.14  & 1.501 \\
  \hline
  15 & 975.3 & 42.72  & 1.822 \\
  \hline
  20& 968.7 & 37.97 & 2.142 \\
  \hline
  40 & 935.3 & 30.16 & 2.846 \\
  \hline
  60 & 891.1 & 26.23 & 2.547 \\
  \hline
  80 & 843.6 & 23.82 & 1.881 \\
   \hline
  100 &789.3 & 21.82 & 1.203 \\
  \end{tabular}}
  \label{tab:fluidparm}
\end{table}

\begin{figure*}[tbh]
\centering
\includegraphics[width=.9\textwidth]{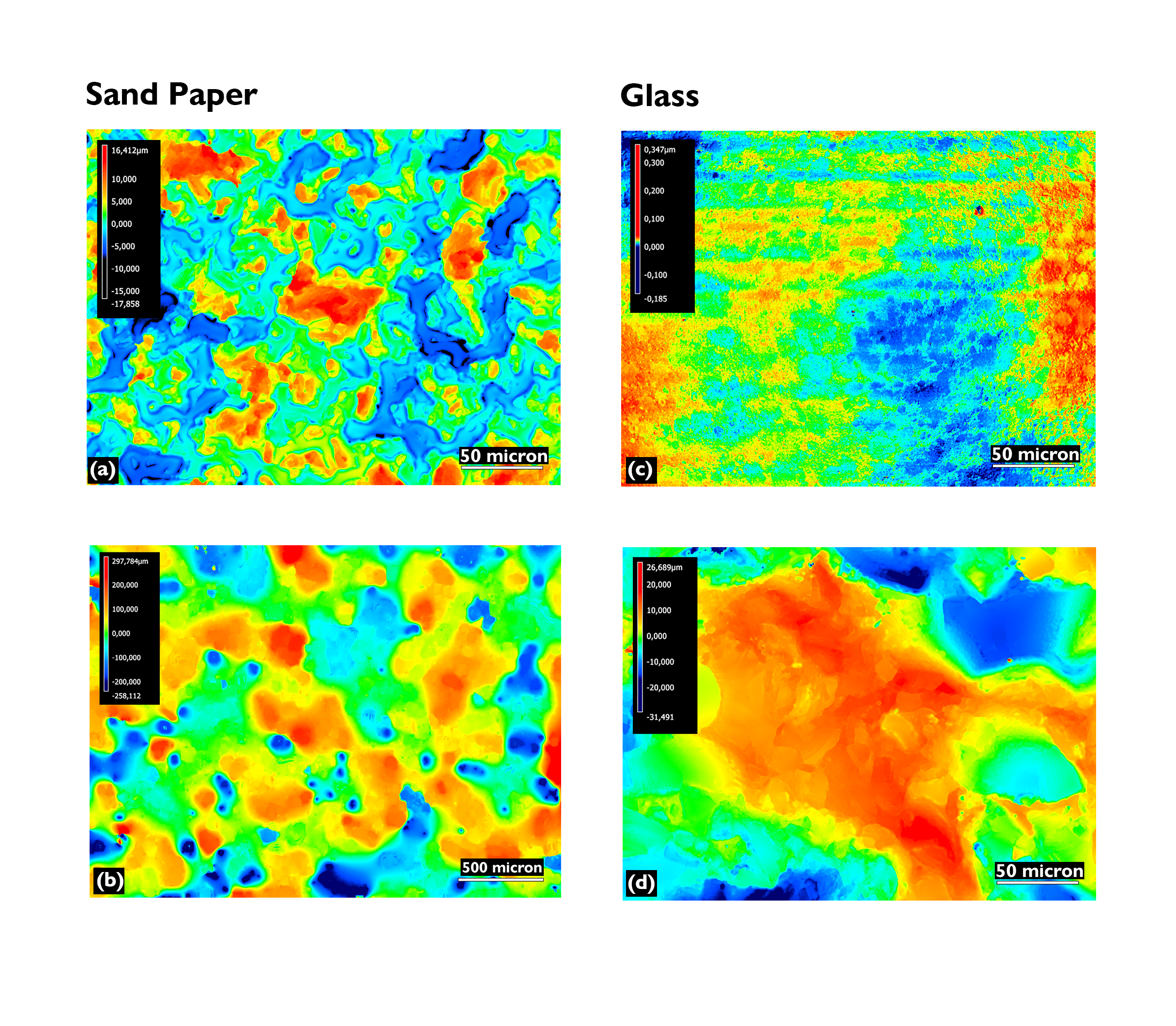}
\caption{Surface roughness profiles of the sand paper surfaces ((a) grit 2000, $R_{rms}= 4.6 \pm 0.3 \hspace{0.1cm} \mu$m ; (b) grit 80, $R_{rms}= 78 \pm 7 \hspace{0.1cm} \mu$m) and glass surfaces ((c) smooth, $R_{rms}= 0.0075 \pm 0.0006 \hspace{0.1cm} \mu$m ; (d) sandblasted , $R_{rms}= 7 \pm 1 \hspace{0.1cm} \mu$m) as measured with a Keyence VK-X1000 laser scanning confocal microscope. The surface roughness profiles shown in (c) and (d) are for untreated glass slides.  The color coding indicates the height or depth of the peaks and valleys with respect with a reference plane of the roughness profile, with the legend shown in the top left of each image.}
\label{fig:key}
\end{figure*}

Droplet splashing was measured by generating droplets from a syringe equipped with a 0.4 mm diameter needle, where the initial droplet diameter $D_0$ varied between 1.8 and 2.2 mm depending on the surface tension of the liquid. The impact of these droplets on a surface was measured using a high speed camera (Phantom Miro M310) at frame rates between 8100 and 32000 frames per second at an optical resolution of around 19 micron per pixel. The height of the needle was increased, thus increasing the impact velocity of the droplet, until at least one satellite droplet was formed during impact, which was defined as the onset of splashing. Measurements were repeated six times and splashing velocities averaged over these six measurements. As experimental error, we use the difference between the measured splashing velocity, and the measured impact velocity at the previous investigated drop height, for which splashing did not occur. All experiments were done at atmospheric conditions, with air density, viscosity and mean free path of the air molecules as reported by Riboux and Gordillo \cite{Riboux2014}.

As liquids, nine different water-ethanol mixtures were used. The density, liquid surface tension and viscosity of these mixtures (given in wt (\%) of ethanol) are shown in Table \ref{tab:fluidparm}. 

As surfaces, sand paper with different grits and (sandblasted) glass slides (Corning Plain Microslides) were used. Their surface roughness was characterised by the root mean square roughness $R_{rms}$, as measured using a Keyence VK-X-1000 laser scanning confocal microscope (Figure \ref{fig:key}). The root mean square roughness of the sand paper lies between $4.6\pm 0.3$ and $78 \pm 7\,\mu$m (Figure \ref{fig:key}a and \ref{fig:key}b). The glass slides were either completely smooth (Figure \ref{fig:key}c) or sandblasted with different sand particle sizes (Figure \ref{fig:key}d), giving a measured range of root mean square roughness between $0.0075 \pm 0.0006$ and $7 \pm 1\,\mu$m. \\  

To investigate the influence of surface wettability, glass slides of different roughnesses were silanised to decrease their surface wettability  \cite{Dey2016}. For all surfaces in this study, the macroscopic contact angle of water was measured for the sand paper and untreated and silanised glass slides using the sessile drop method \cite{Bachmann2000,Bachmann2000b}.\\

\section{Results and Discussion}

\subsection{Splashing Mechanism}

\begin{figure*}[htb]
\centering
\includegraphics[width=.95\textwidth]{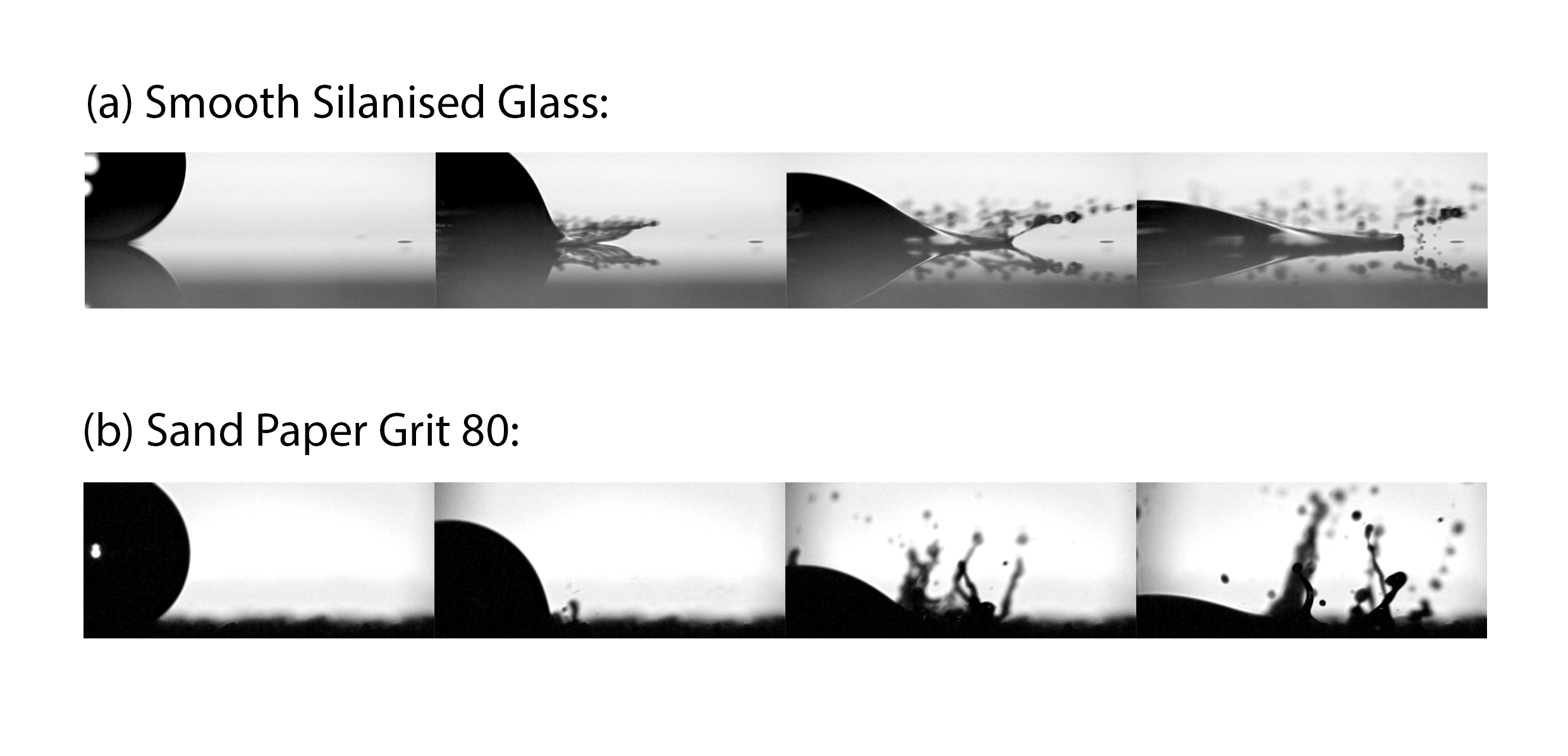}
\caption{High speed images of an ethanol droplet impacting (a) a smooth silanised glass slide ($R_{rms} = 0.03 \pm 0.03\,\mu$m) and (b) grit 80 sand paper ($R_{rms} = 78 \pm 7\,\mu$m) (b) at the same impact velocity ($v > v_{sp}$). For the smooth surface, satellite droplets are formed from the rim of the lifted sheet of liquid (corona splash), whereas for rough surfaces the satellite droplets are formed from finger-like structures (prompt splash).}
\label{fig:corvsprompt}
\end{figure*}

Figure \ref{fig:corvsprompt} shows splashing of an ethanol droplet impacting a smooth silanised glass slide (Figure \ref{fig:corvsprompt}a) and the sand paper with the largest grit size (Figure \ref{fig:corvsprompt}b). Comparing the two image sequences shows the difference in splashing mechanism for smooth and rough surfaces. For smooth surfaces (Figure \ref{fig:corvsprompt}a), a corona splash occurs, where the satellite droplets detach at the edge of the lifted liquid lamella. Corona splashing on smooth surfaces can be described by the splashing model of Riboux and Gordillo \cite{Riboux2014}, who predict the splashing velocity of droplet splashing on smooth surfaces by also considering the properties of the surrounding air. For rough surfaces, a different splashing mechanism is observed. Instead of the formation of a lamella that radially spreads outwards, liquid fingers seem to form that subsequently break up into smaller droplets (Figure \ref{fig:corvsprompt}b), which is known as prompt splashing. Therefore, these images suggests that a transition takes place in the splashing mechanism when the surface roughness is increased.\\

\begin{figure*}[t]
	\centering
	\includegraphics[width= 0.6\textwidth]{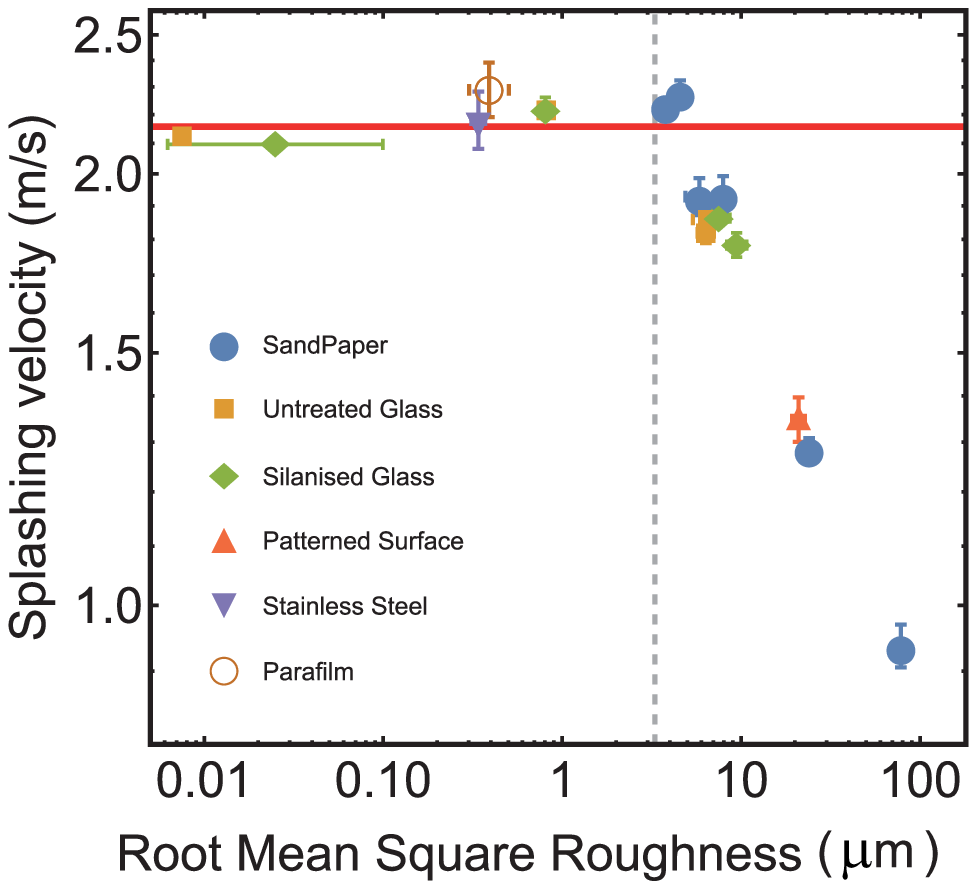}
	\caption{ Measured splashing velocity as function of the surface roughness plotted on log-log scale for ethanol impacting sand paper (blue circles), and untreated (yellow squares) and silanised (green diamonds) glass slides. The red line gives the theoretical prediction for corona splashing \cite{Riboux2014}. The splashing velocity measurements for ethanol impacting stainless steel (purple downward triangle) and parafilm (open brown circle) from the study of de Goede \textit{et al.} \cite{deGoede2017} are added for comparison. The vertical grey dashed line shows the transition roughness calculated with Eq. \eqref{eq:transrough}}
	\label{fig:splashvel}
\end{figure*}

Figure \ref{fig:splashvel} shows the splashing velocity of ethanol as a function of the surface roughness. For ethanol, the splashing velocity seems to be constant up until a certain root mean square roughness ($\sim 3-4\,\mu$m), after which the splashing velocity decreases with increasing roughness. Comparing the data with the splashing velocity for stainless steel (downward purple triangle) and parafilm (open brown circle) from de Goede \textit{et al.} \cite{deGoede2017} shows a good agreement. Furthermore, no significant difference in the roughness dependence of the splashing velocity is observed between the sand paper (blue circles), untreated (yellow squares) and silanised (green diamonds) glass slides. As ethanol has a low liquid surface tension and therefore fully wet all surfaces ($\theta = 0^\circ$), this does not rule out a surface wettability dependence.\\ 

Comparing the measured splashing velocity with the prediction by Riboux and Gordillo for corona splashing \cite{Riboux2014} (red line in Figure \ref{fig:splashvel}) shows that for surface roughnesses below $4\,\mu$m, the splashing velocity is predicted very well by this model. The change in the dependence of the splashing velocity on the surface roughness happens simultaneously with the change in the splashing mechanism from corona to prompt splashing (Figure \ref{fig:corvsprompt}). When the roughness exceeds $4\,\mu$m, it starts to affect the liquid flow on the surface and breaks up the lamella into the finger-like structures observed in Figure \ref{fig:corvsprompt}b, which also influences the splashing velocity.\\

\begin{figure*}[tbh]
\centering
\subfloat[]{\includegraphics[width =.48\textwidth]{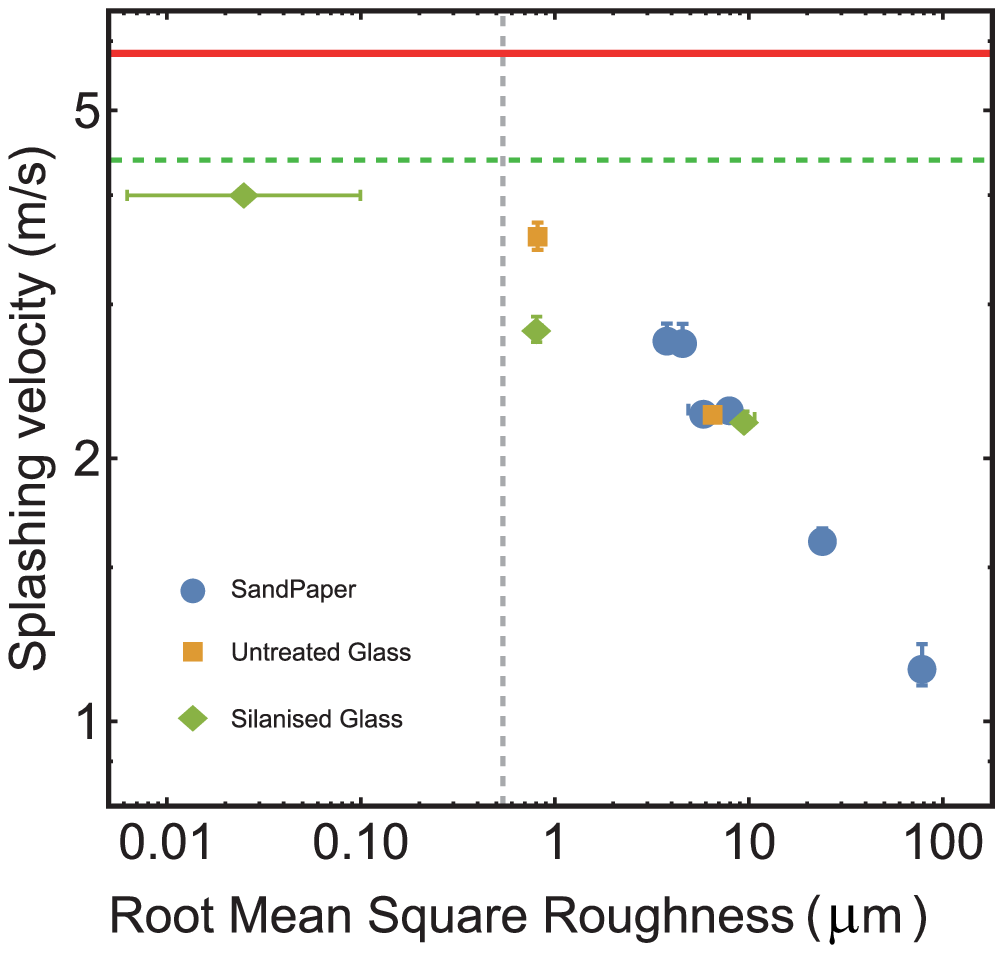}} \hfill
\subfloat[]{\includegraphics[width=.49\textwidth]{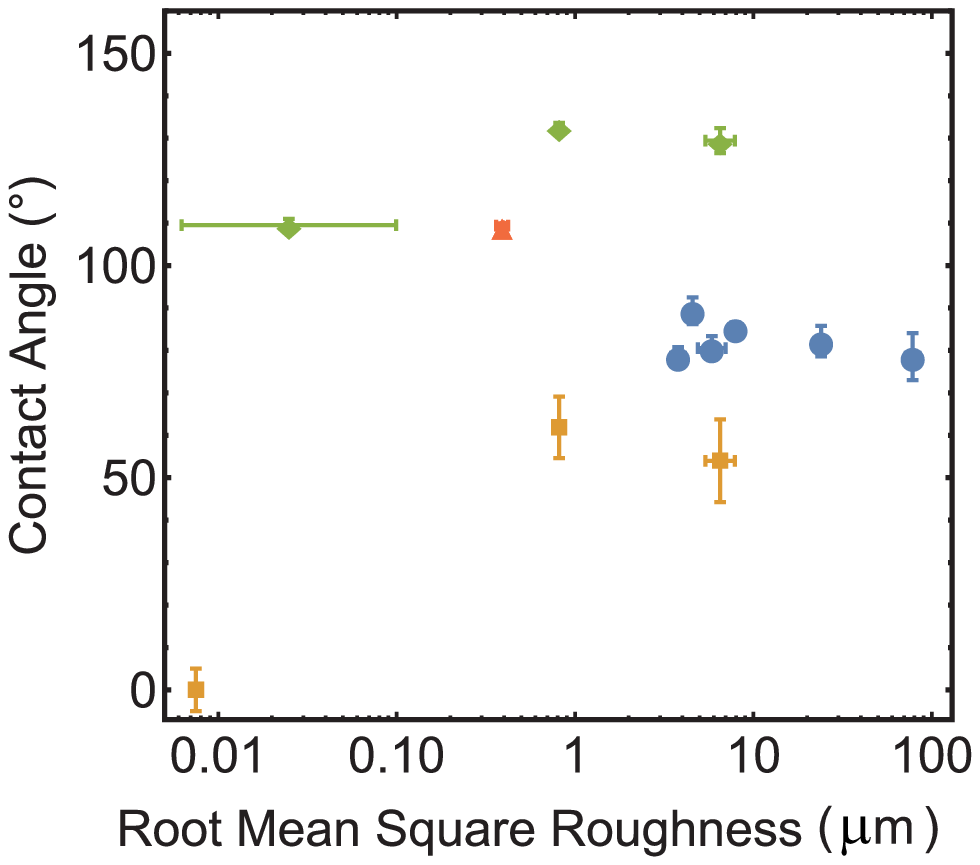}}
\caption{(a) Measured splashing velocity of water as function of the surface roughness on sand paper (blue circles), and untreated (yellow squares) and silanised (green diamonds) glass slides plotted on log-log scale. The red line gives the theoretical prediction for corona splashing \cite{Riboux2014}. No splashing was observed for droplet impact on the smooth untreated glass surface, where the highest experimentally measured impact velocity is denoted by the horizontal, green dashed line.
The vertical grey dashed line shows the transition roughness calculated with Eq. \eqref{eq:transrough}. (b) Measured contact angles of water as a function of the root mean square roughness of the sand paper (blue circles), untreated (yellow squares) and silanised glass slides (green diamonds). The contact angle of parafilm from \cite{deGoede2017} (red square) is added to this graph as well.} 
\label{fig:contactangles}
\end{figure*}

\subsection{Surface Wettability}

In order to investigate the influence of wettability, we switch to a higher surface tension liquid, i.e. water. Figure \ref{fig:contactangles} shows the splashing velocity of water droplets as a function of surface roughness for various types surfaces. All measured splashing velocities are significantly lower compared to the prediction of the splashing model of Riboux and Gordillo. In fact, this prediction is higher than the highest impact velocity experimentally accessible to us ($\sim 4$ m/s). As a result of this upper bound, no droplet splashing is observed for pure water on the smooth untreated glass slide. \\ 

We find that the splashing velocity of water on the silanised glass slides (green diamonds) is significantly lower compared to the water splashing velocity on the untreated glass slides (yellow squares). The measurements show that this difference in splashing velocity between the two types of glass slides decreases with increasing roughness, where it becomes comparable to the measured splashing velocity for water on sand paper. Comparing the measured contact angle of these three surfaces (Figure \ref{fig:contactangles}b) shows that the contact angles ranges widely from fully wetting (the smooth untreated glass slide) to around $135^\circ$ for the rough silanised glass slides. The contact angle measurements of both the untreated and silanised glass slides show a relation between surface roughness and surface wettability: By increasing the surface roughness of the glass slides the contact angle of the surface also increases. It is possible that is due to a transition from the liquid being in contact with the whole surface (Wenzel state) to a state where the `mountains' of the surface roughness act like a fakir bed on which the droplet rests (Cassie-Baxter state). However, whether the Wenzel to Cassie-Baxter transition indeed takes places cannot be determined with the measurements shown here and is beyond the scope of this study.\\

These results are in agreement with the recent study of Quetzeri \emph{et al.} \cite{Quetzeri2019}, who showed that the surface wettability starts to have an influence when the surface becomes hydrophobic ($\theta >90^\circ $). This effect of surface wettability was not observed in the study of de Goede \textit{et al.} \cite{deGoede2017}, as the highest measured contact angle there was on the order of $90^\circ $. For the liquid-surface combination (5 wt$\%$ ethanol mixture impacting parafilm) used in that study, a lower splashing velocity is observed (Figure 3 of \cite{deGoede2017}), but the difference is within the error margin originating from the used setup. We conjecture that low surface wettability influences droplet splashing by changing the liquid droplet's shape at the contact line, which most likely changes the interaction between the surrounding air and the liquid when splashing occurs. However, with the measurements presented here, a definite conclusion cannot be given on how exactly surface wettability influences droplet splashing in the low wettability regime. For the remainder of this study, we consider liquid-surface combinations that correspond to good wettability to minimise the influence of low surface wettability on droplet splashing. \\

\subsection{Splashing Scaling Models}
 
\begin{figure*}[tb]
\centering
\subfloat[]{\includegraphics[width=.5\textwidth]{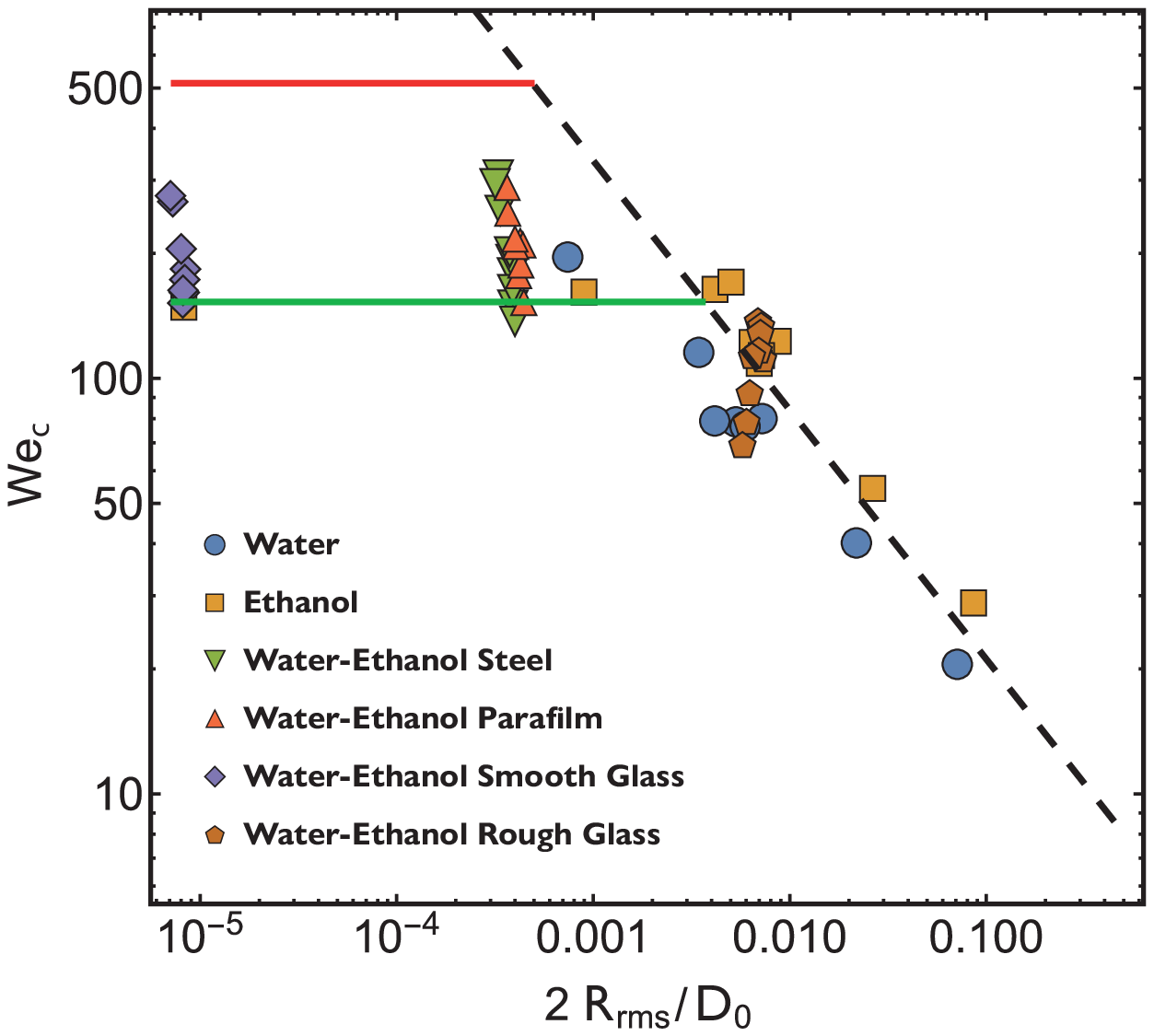}}
\subfloat[]{\includegraphics[width=.5\textwidth]{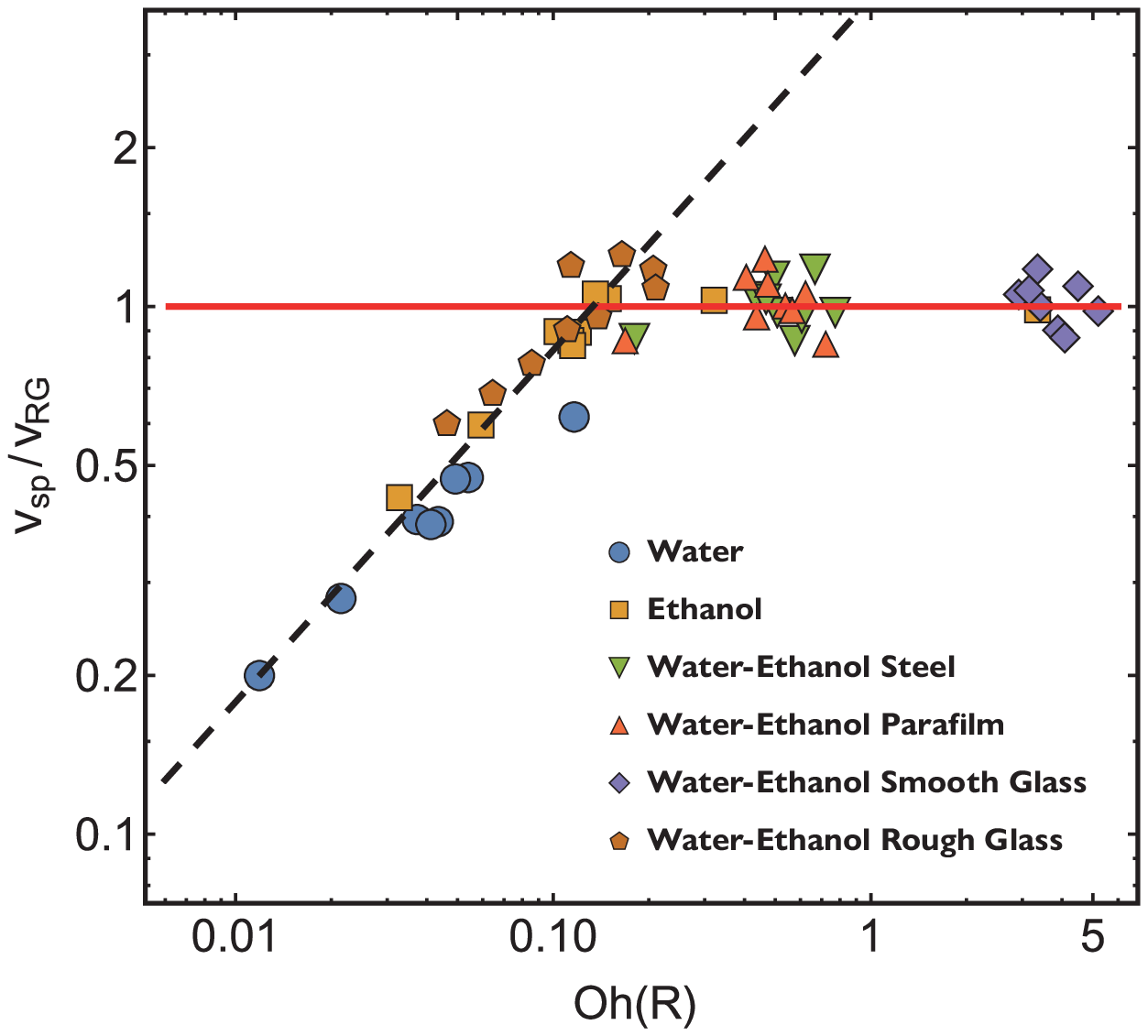}}
\caption{(a) Calculated critical Weber number as a function of $2 R_{rms} / D_0$ for the water (blue circles) and ethanol (yellow squares) measurements shown in Figures \ref{fig:splashvel}a and \ref{fig:splashvel}b, together with measurements of water-ethanol mixtures on rough glass (brown pentagons). Data for water-ethanol mixtures on stainless steel (downward green triangles), parafilm (upward red triangles) and smooth glass (purple diamonds) from de Goede \textit{et al.} \cite{deGoede2017} are included. The red and green solid lines are the critical Weber numbers calculated using the splashing model of Riboux and Gordillo \cite{Riboux2014} for water and ethanol, respectively. The black dashed line is given by $We_c \geq K_2 \left(\cos \theta_0\right) \left( 2 R_{rms} / D_0 \right)^{-3/5}$ (Eq. \ref{eq:wegor}). (b) Ratio $v_{sp}/v_{RG}$ as a function of the Ohnesorge number with $R_{rms}$ as characteristic length scale $Oh(R_{rms}0)$ for the water and ethanol measurements on different surfaces (blue circles and yellow squares, respectively) and water-ethanol mixtures on a rough glass slide (brown pentagons). Data for water-ethanol mixtures on stainless steel (green triangles), smooth glass (purple diamonds) and parafilm (red triangles) from de Goede \textit{et al.} \cite{deGoede2017} are included. The red line gives the prediction of the splashing model of Riboux and Gordillo, and the black line is the best fit of $Oh^{2/3}$}
\label{fig:wecscale}
\end{figure*}

Our observation that the splashing velocity depends on surface roughness above a critical roughness corresponding to the transition from corona to prompt splashing suggests that the finger like break up of the droplet is most likely caused by the droplet pushing itself into the 'valleys' of the surface roughness, and then ejecting a liquid finger when it encounters a 'mountain'. Garc{\'i}a-Geijo  and collaborators \cite{Quintero2020} used this idea to explain prompt splashing in the low Ohnesorge number limit: When the surface roughness becomes comparable to the the thickness of the spreading lamella $H_t$ on top of the surface, the droplet can push itself into the surface roughness if the impact velocity is high enough and the droplet's inertial forces are strong enough to overcome the resisting capillary pressure of the surface. In this case, the spreading lamella of the droplet breaks up and results in the change of splashing mechanism from corona to prompt splashing. With this idea, Garc{\'i}a-Geijo  \textit{et al.} were able to relate the critical Weber number ($We_c = \frac{\rho D_0 v_{sp}}{2 \sigma}$) of the droplet at the moment of splashing to a length scale ratio of the surface roughness and the initial radius ($D_0 / 2$) of the droplet:

\begin{equation}
We_c \geq K_2 \left( 8 \cos \theta_0 \right)^{3/5} \left(\frac{2 R_{rms}}{D_0}\right)^{-3/5}
\label{eq:wegor}
\end{equation}  

Where $K_2$ is a order unity constant and $\theta_0$ the static contact angle between the liquid and smooth surface. In Figure \ref{fig:wecscale}a we calculate the critical Weber number and length scale ratio corresponding to our experimental results for water (blue circles) and ethanol (yellow squares), revealing that they roughly collapse onto the same curve. Similar to the direct measurements of the splashing velocity, the experimentally determined critical Weber number remains independent of the length scale ratio up until $ \frac{2 R_{rms}}{D_0} \approx 0.004$, after which both the water and ethanol data again decrease with increasing $2 R_{rms}/D_0$. Fitting Eq. \eqref{eq:wegor} using the least square method to the critical Weber number above the critical length scale ratio shows that the model of Garc{\'i}a-Geijo  \textit{et al.} predicts the power law decrease in the critical Weber number relatively well (black dashed line Figure \ref{fig:splashvel}). As we do not know $\theta_0$ for the sand paper surfaces, $K_2 \left(8 \cos \theta_0\right)^{3/5}$ is used as the fitting parameter for both the glass and sand paper surfaces in this study, for which we find an average value of $5.2 \pm 0.3$.\\

The splashing data of water-ethanol mixtures on glass (purple diamonds), stainless steel (green downward triangles) and parafilm (red upward triangles) surfaces from the study of de Goede \textit{et al.} \cite{deGoede2017} were added to Figure \ref{fig:wecscale}a. These measurements do not seem to fully collapse on the same curve as the splashing velocity measurements from the present study: The critical Weber number seems to still increase with increasing surface tension. Calculating the critical Weber number using the splashing model of Riboux and Gordillo, the value $We_c$ for water (red line Figure \ref{fig:wecscale}a) is much higher compared to the value of ethanol (green line Figure \ref{fig:wecscale}a), suggesting that the differences among $We_c$ values for the water-ethanol mixtures therefore are caused by differences in the surface tension. We find similar differences when investigating water-ethanol mixtures on a rough glass slide. \\

The experimental results also show that the surface roughness at which the splashing mechanism transition takes place is dependent on the surface tension of the liquid. This is explained by the scaling model of Garc{\'i}a-Geijo  \textit{et al.} \cite{Quintero2020}, as a higher liquid surface tension increases the capillary pressure that the droplet's inertial forces need to overcome. It is possible to determine this `transition roughness' $R_c$ for water and ethanol by calculating the roughness at the intersection between the best fit of Eq. \eqref{eq:wegor}, and the critical Weber number determined from the splashing model of smooth surfaces \cite{Riboux2014}:

\begin{align}
R_{c} = \frac{D_0}{2}\left(\frac{We_c}{5.3}\right)^{-5/3}
\label{eq:transrough}
\end{align}  

For ethanol, this calculated roughness is $3.3 \hspace{0.1cm} \mu$m and agrees very well with the moment the measured splashing velocity of ethanol starts to decrease with increasing surface roughness (grey dashed line in Figure \ref{fig:splashvel}). Furthermore, the transition roughness for water ($0.53 \hspace{0.1cm} \mu$m) also seems to agree with the experiments, as this calculated roughness is lower than the lowest surface roughness for which splashing of water was measured experimentally. The agreement between theory and experiments could indicate that the capillary pressure caused by the surface roughness plays a role in how a droplet splashes. \\  

Although the critical Weber number scaling model of Garc{\'i}a-Geijo  \textit{et al.} works relatively well, it does not fully collapse the data onto a single curve as we have shown above. The problem with using the Weber number is that it neglects the liquid viscosity, which has a significant influence on droplet splashing, such as in the splashing model of Riboux and Gordillo used here. To fully understand droplet splashing on rough surfaces, the liquid viscosity also has to be incorporated.\\ 

To do this, we define a velocity ratio given by the measured splashing velocity and the theoretical prediction of the splashing model of Riboux and Gordillo \cite{Riboux2014}. As was shown that this model accurately predicts corona splashing when the surface is smooth enough, the velocity ratio will be equal to one for corona splashing and decrease when the surface roughness starts to change the splashing mechanism into prompt splashing. To incorporate the influence of the fluid parameters (liquid density, surface tension and viscosity) we use the Ohnesorge number ($Oh = \eta / \sqrt{\rho \sigma L}$), where $\eta$ is the visocosity, $\rho$ the density, $\sigma$ the surface tension, and $L$ the typical length scale of the system. As prompt splashing of the liquid is likely caused by the surface roughness, we use the root mean square roughness $R_{rms}$ as the length scale. When the velocity ratio and the Ohnesorge number are determined from the same experimental data that were also used to test the scaling model of Garc{\'i}a-Geijo  \textit{et al.}, we find that all the experimental data collapse onto a single curve (Figure \ref{fig:wecscale}b). Again, a transition between two regimes can be found, where the velocity ratio increases with Ohnesorge number until it reaches unity (red line) around $Oh \approx 0.15$, above which the velocity ratio is independent of the Ohnesorge number. Fitting the experimental velocity ratio using the least squares method shows that the velocity ratio increases with $Oh^{2/3}$, or $R_{rms}^{-1/3}$, until it reaches unity. \\

While our Ohnesorge scaling model is an excellent fit to the experimental data, it lacks theoretical support. It is also important to note that the influence of the surrounding air has not been investigated in this study. The surrounding air (pressure) not only has an influence on corona splashing but also on prompt splashing, which can be influenced by changing the atmospheric pressure \cite{Latka2012}.

\section{Conclusions}

In this study, we investigated the influence of the surface roughness on droplet splashing. By changing the root mean square roughness of the impacted surface, we showed that the droplet splashing velocity is only affected when the droplet roughness is large enough to change to disrupt the spreading droplet lamella and change the droplet splashing mechanism from corona to prompt splashing. Finally, using Weber and Ohnesorge number scaling models, we have shown that the measured splashing velocity for both the water and ethanol on different surface roughness and water-ethanol mixtures collapse on a single curve, showing that the splashing velocity on rough surfaces scales with $Oh^{2/3}$ at atmospheric pressures.\\

These results suggest that it is possible, using the roughness of the surface, to determine whether corona splashing or prompt splashing will occur, and whether the splashing model for smooth surfaces can be applied. For the prompt splashing regime, we show that an Ohnesorge scaling with the surface roughness as typical length scale can be used to predict deviations from the splashing model by Riboux and Gordillo. While not yet theoretically underpinned, the Ohnesorge model is a first step in the direction of a scaling model that takes surface tension into account.

\bibliography{biblio}

\end{document}